\documentclass[superscriptaddress,twocolumn,10pt,prl,floatfix,nofootinbib,aps]{revtex4-2}

\usepackage{amsmath,amssymb}
\usepackage{graphicx}
\usepackage{siunitx}
\usepackage{lipsum}
\usepackage{xcolor}
\usepackage{booktabs}
\usepackage[hidelinks]{hyperref}
\usepackage{xurl}
\usepackage{siunitx}
\usepackage{multirow}
\usepackage{caption}
\usepackage{subcaption}
\usepackage{ragged2e}

\usepackage{natbib}
\usepackage{braket}

\newcommand{\imec}{Imec, Leuven, Belgium}
\newcommand{\diraq}{Diraq, Sydney, New South Wales, Australia}
\newcommand{\esat}{Department of Electrical Engineering (ESAT), KU Leuven, Leuven, Belgium}
\newcommand{\phys}{Department of Physics and Astronomy, KU Leuven, Leuven, Belgium}
\newcommand{\ghent}{Department of Solid State Sciences, Ghent University, Ghent, Belgium}
\newcommand{\unsw}{School of Electrical Engineering and Telecommunications,\\University of New South Wales, Sydney, New South Wales, Australia}
\newcommand{\proximus}{Proximus Chair in Quantum Science and Technology, Department of Electrical Engineering 
(ESAT), KU Leuven, Leuven, Belgium}

\newcommand{\equalcontribmark}{\textsuperscript{+}}
\newcommand{\equalcontribtext}{These authors contributed equally to this work.}

\DeclareCaptionLabelSeparator{line}{\space\textbar\space}

\begin{document}

\captionsetup[figure]{name={\bf{Figure}},labelsep=line,justification=raggedright,font=small,singlelinecheck=false}
\captionsetup[table]{name={\bf{Table}},labelsep=line,justification=raggedright,font=small,singlelinecheck=false}

\title{SiMOS quantum-dot spin qubits enabled by extreme-ultraviolet lithography}

\author{\textbf{Thomas Van Caekenberghe}\equalcontribmark}
\email{thomas.vancaekenberghe@imec.be}
\affiliation{\imec}
\affiliation{\esat}

\author{\textbf{Paul Steinacker}\equalcontribmark}
\email{paul.s@diraq.com}
\affiliation{\diraq}
\affiliation{\unsw}

\author{Bart Raes}
\author{Sofie Beyne}
\author{Clement Godfrin}
\author{Jacques Van Damme}
\author{Sylvain Baudot}
\affiliation{\imec}

\author{Arne Loenders}
\affiliation{\imec}
\affiliation{\esat}

\author{Gulzat Jaliel}
\author{Stefan Kubicek}
\author{Johan De Backer}
\author{Yannick Hermans}
\author{Sugandha Sharma}
\author{Shuchi Kaushik}
\author{Yuchao Jiang}
\author{Yosuke Shimura}
\affiliation{\imec}
\author{Roger Loo}
\affiliation{\imec}
\affiliation{\ghent}

\author{Vukan Levajac}
\affiliation{\imec}
\affiliation{\phys}
\author{Kristof Moors}
\author{George Simion}
\affiliation{\imec}

\author{Florian K. Unseld}
\author{Ensar Vahapoglu}
\author{Ajit Dash}
\author{Tuomo Tanttu}
\affiliation{\diraq}
\affiliation{\unsw}
\author{Chris C. Escott}
\affiliation{\diraq}
\author{Chih Hwan Yang}
\affiliation{\diraq}
\affiliation{\unsw}
\author{Andre Saraiva}
\affiliation{\diraq}
\author{Arne Laucht}
\author{Wee Han Lim}
\affiliation{\diraq}
\affiliation{\unsw}

\author{Nard Dumoulin Stuyck}
\affiliation{\diraq}
\affiliation{\unsw}

\author{Massimo Mongillo}
\author{Danny Wan}
\affiliation{\imec}

\author{Andrew S. Dzurak}
\affiliation{\diraq}
\affiliation{\unsw}

\author{Kristiaan De Greve}
\email{kristiaan.degreve@imec.be\\}
\affiliation{\imec}
\affiliation{\proximus}

\date{\today}

\begin{abstract}
{The realization of large-scale silicon quantum processors requires spin qubits compatible with advanced semiconductor manufacturing technologies, demanding lithographic processes that combine nanometer-scale precision with exceptional uniformity. Although the highest-performing silicon spin qubits demonstrated to date have relied on electron-beam (e-beam) lithography, its serial exposure process limits reproducibility studies and wafer-scale fabrication. Here, we demonstrate high-performance silicon metal-oxide-semiconductor (SiMOS) spin qubits fabricated using extreme-ultraviolet (EUV) lithography in a \SI{300}{\milli\meter} semiconductor pilot line. We report wafer-scale quantum-dot uniformity metrics, including \SI{100}{\percent} room-temperature gate-to-gate leakage yield and sub-nanometer control of critical gate dimensions. We characterize four double-dot systems realized in two triple-quantum-dot devices. Gate set tomography (GST) reveals consistently high fidelities across all four systems, with values up to \SI{99.8}{\percent} for SPAM, \SI{99.9}{\percent} for single-qubit gates, and \SI{99.1}{\percent} for two-qubit gates. The devices exhibit highly reproducible exchange turn-on characteristics of \SI{10}{}–-\SI{13}{dec}$\mathrm{V}^{-1}$, indicating high fabrication uniformity enabled by EUV patterning. These results establish EUV lithography as a viable manufacturing technology for quantum processors based on high-fidelity SiMOS spin qubits.}
\end{abstract}

\maketitle

\renewcommand{\thefootnote}{+}
\footnotetext{\equalcontribtext}

\noindent
The prospect of a universal quantum computer built from silicon spin qubits has gained considerable momentum in recent years, driven by their nanometer-scale footprint, long coherence times, high-fidelity operations, classical control electronics integration, and compatibility with complementary metal-oxide-semiconductor (CMOS) manufacturing~\cite{loss_quantum_1998,zwanenburg_silicon_2013,veldhorst_addressable_2014,maurand_cmos_2016,xue_cmos-based_2021,ruffino_cryo-cmos_2021,pauka_cryogenic_2021,stano_review_2022,neyens_probing_2024,elsayed_low_2024,george_12-spin-qubit_2025,bartee_spin-qubit_2025,members_of_the_hrl_quantum_team_digitally_2026,dumoulin_stuyck_cmos_2026}. This compatibility is widely regarded as a key advantage for scaling beyond the relatively small arrays demonstrated to date and ultimately realizing fault-tolerant quantum computing~\cite{veldhorst_silicon_2017,vandersypen_interfacing_2017,campbell_roads_2017,zwerver_qubits_2022}. The path toward large-scale manufactured qubit chips requires lithographic processes that simultaneously provide the dimensional precision and reproducibility needed for high-fidelity qubit operation, while enabling wafer-scale fabrication for statistical process optimization.

To date, the highest-performance silicon spin qubit devices have been defined using electron-beam (e-beam) lithography~\cite{yoneda_quantum-dot_2018, yang_silicon_2019,noiri_fast_2022,xue_quantum_2022,mills_two-qubit_2022,tanttu_assessment_2024,steinacker_bell_2025,steinacker_industry-compatible_2025}. E-beam writing offers high resolution and flexibility, making it the tool of choice for prototyping nanoscale gate architectures. However, the serial nature of pattern writing requires complex layouts to be assembled from many individual exposures, making the process susceptible to stitching errors and contributing to limitations in overlay accuracy between successive patterning layers~\cite{vieu_electron_2000,lawrie_quantum_2020}. Even nanometer-scale variations in gate placement and geometry translate directly into inter- and intra-device variability, increasing calibration overhead and limiting the scalability of large spin-qubit arrays~\cite{borsoi_shared_2024,tosato_crossbar_2026,li_tri-linear_2026}. In addition, the serial nature of e-beam writing causes exposure times to increase linearly with the number and size of patterned structures, rendering full-wafer fabrication impractical.

In contrast, semiconductor manufacturing relies almost exclusively on optical lithography owing to its throughput, excellent overlay control, and process maturity. Extreme-ultraviolet (EUV) lithography extends optical lithography to the \SI{13.5}{\nano\meter} wavelength regime, enabling the patterning of tight-pitch gate geometries required for silicon spin-qubit devices~\cite{radamson_cmos_2024,giannopoulos_extreme_2024}. EUV lithography transfers an entire reticle field in a single exposure, providing high-throughput fabrication with overlay precision below \SI{2}{\nano\meter} through advanced alignment schemes~\cite{thijssen_cross-platform_2018}. The exposed field size is substantially larger than the dimensions of present-day spin-qubit chips, allowing complex quantum-device layouts to be fabricated without stitching between exposure fields. As a result, identical dies can be replicated across an entire \SI{300}{\milli\meter} wafer with excellent dimensional control and overlay accuracy. Moreover, adopting EUV lithography aligns qubit fabrication with the broader semiconductor manufacturing ecosystem, providing access to mature process control, metrology, and integration capabilities refined over decades of industrial development.

Although EUV patterning has been demonstrated for silicon spin qubits in Si/SiGe heterostructures~\cite{george_12-spin-qubit_2025}, it remains unclear whether the technology can satisfy the more stringent dimensional requirements of SiMOS quantum dots. Whereas typical Si/SiGe devices employ dot-to-dot pitches on the order of 120--\SI{140}{\nano\meter}, SiMOS quantum dots generally require pitches below \SI{100}{\nano\meter} to achieve few-electron confinement and sufficiently strong exchange interaction~\cite{lawrie_quantum_2020,scappucci_crystalline_2021}. This aggressive scaling places stringent demands on lithographic resolution, dimensional control, and layer-to-layer overlay, extending EUV lithography into a particularly demanding regime for gate-defined quantum devices.

In this work, we demonstrate excellent uniformity of SiMOS quantum dot devices and report the first high-fidelity SiMOS spin qubits fabricated using EUV lithography. We first assess the fabrication uniformity of a triple-quantum-dot design with a targeted dot-to-dot pitch of \SI{80}{\nano\meter} across a \SI{300}{\milli\meter} wafer, demonstrating \SI{100}{\percent} room-temperature gate-to-gate leakage yield together with highly reproducible gate dimensions, oxide thickness, and inter-layer overlay. We then perform an in-depth characterization at sub-Kelvin temperatures of two more aggressively scaled \SI{60}{\nano\meter}-pitch triple-quantum-dot devices fabricated on the same wafer. Operating each device as two independent double-dot systems, we benchmark their performance using gate set tomography (GST)~\cite{blume-kohout_demonstration_2017,nielsen_gate_2021}. Across all four double-dot systems, we demonstrate the full set of fundamental qubit operations, including single- and two-qubit gates, initialization, and readout, with consistently low error rates approaching the fault-tolerance threshold commonly associated with the surface error correction code~\cite{fowler_surface_2012}. The observed reproducibility across double-dot systems and devices is further reflected in the reproducible exchange turn-on of 10--\SI{13}{dec}V$^{-1}$. These results establish EUV lithography as a viable manufacturing technology for high-fidelity SiMOS spin qubits, bridging the gap between prototype demonstrations and industrial semiconductor manufacturing.

\section*{Device fabrication}
\begin{figure}[b]
\includegraphics[width=\columnwidth]{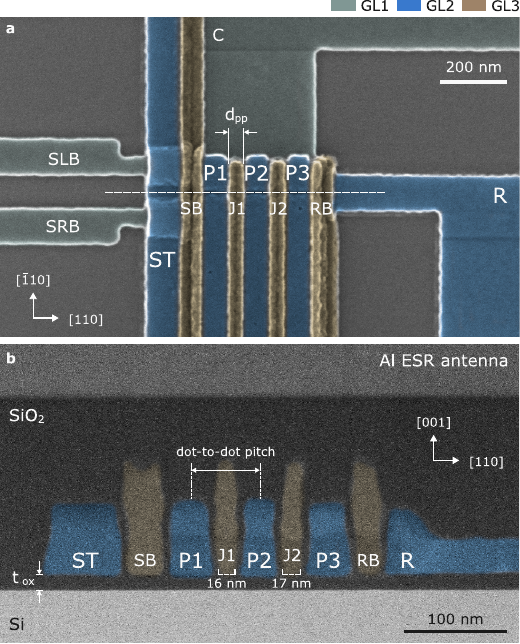}
\caption{\label{fig:m1}\textbf{Triple-quantum-dot device.} \textbf{a,} False-colored top-view scanning electron micrograph (SEM) of a triple-quantum-dot device. The device comprises three overlapping polysilicon gate layers and consists of two confinement gates (C) at the top and bottom (not visible) of the dots in gate layer 1 (GL1), three plunger gates (P1, P2, P3) in gate layer 2 (GL2), and two exchange gates (J1, J2) in gate layer 3 (GL3). A single-electron transistor positioned to the left of the quantum-dot array is used for charge sensing. The SEM image shows a representative device with an \SI{80}{\nano\meter} dot-to-dot pitch. \textbf{b,} False-colored cross-sectional transmission electron microscopy (TEM) image of device A, acquired along a cross-section corresponding to the dashed line indicated in \textbf{a}. Device A implements the same gate architecture with a reduced dot-to-dot pitch of \SI{60}{\nano\meter}. The widths of exchange gates J1 and J2, measured at the Si/SiO$_2$ interface, are \SI{16}{\nano\meter} and \SI{17}{\nano\meter}, respectively.}
\end{figure}

\noindent The qubit devices were designed and fabricated at imec using a \SI{300}{\milli \meter} spin-qubit process flow~\cite{beyne_300mm_2025, de_backer_advances_2026}. Fabrication was carried out on high-resistivity silicon substrates with a \SI{100}{\nano\meter} epitaxial layer of isotopically enriched silicon (400 ppm \textsuperscript{29}Si). An \SI{8}{\nano\meter} gate oxide was grown by high-temperature dry oxidation. The gate layers were patterned using single-exposure 0.33 NA EUV lithography and a state-of-the-art metal--oxide resist (MOR), enabling precise and uniform definition of critical features. For each gate layer, \SI{30}{\nano\meter} of highly doped polysilicon was deposited and subsequently patterned via a subtractive etch process. Adjacent gate layers were electrically isolated by \SI{5}{\nano\meter} of high-temperature oxide (HTO). Following completion of the gate stack, a high-temperature anneal was performed to activate the polysilicon gates. Thereafter, contacts, control modules, and passivation layers were fabricated using deep-ultraviolet (DUV) lithography, employing the process flow previously reported for e-beam-defined devices~\cite{li_flexible_2020}.

The wafer incorporates a range of device variants spanning different quantum-dot array geometries and dot-to-dot pitches to probe both process and device limitations. To evaluate the scaling limits of EUV lithography for SiMOS qubits, designs with dot-to-dot pitches ranging from \SI{60}{\nano\meter} to \SI{100}{\nano\meter} were included. The devices considered in this work comprise a triple quantum dot coupled to a proximal single-electron transistor (SET) for readout. Figure~\ref{fig:m1}a shows a representative device with an \SI{80}{\nano\meter} dot-to-dot pitch, corresponding to the center of the process window and the nominal target dimension. In all devices, electrons confined beneath plunger gates P1, P2, and P3 are pairwise coupled via exchange gates J1 and J2, while surrounding confinement (C) and barrier (SB, RB) gates define the electrostatic potential landscape. Electrons are loaded through a reservoir gate (R) overlapping an n\textsuperscript{++} ohmic implant.

Figure~\ref{fig:m1}b shows a cross-sectional TEM image of the gate-stack geometry at the Si/SiO$_2$ interface for a more aggressively scaled \SI{60}{\nano\meter}-pitch device. Substantial process development was undertaken to reduce charge noise~\cite{elsayed_low_2024}, suppress gate-to-gate leakage, and to optimize etch conditions for uniform oxide thickness (Extended Data Fig.~\ref{fig:e1}), thereby ensuring consistent lever arms~\cite{loenders_understanding_2026}; see Ref.~\cite{beyne_300mm_2025} for details. The oxide beneath GL2 and GL3 is approximately \SI{12}{\nano\meter} thick, corresponding to the combined thickness of the thermally grown oxide and the HTO deposited after gate layer 1 (GL1). The plunger gates are approximately \SI{30}{\nano\meter} wide at the interface. Combined with the \SI{60}{\nano\meter} dot-to-dot pitch and \SI{5}{\nano\meter} inter-layer oxide, this geometry yields exchange gates narrower than \SI{20}{\nano\meter}, enabling precise control of the exchange interaction between neighboring dots.

\begin{figure}[t]
\includegraphics[width=\columnwidth]{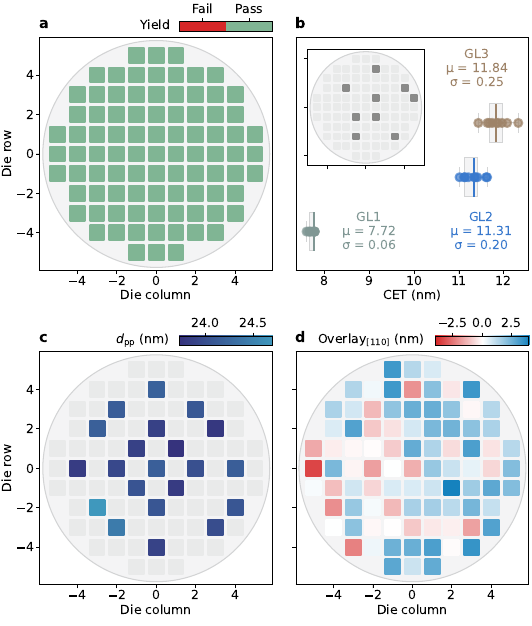}
\caption{\label{fig:m2}\textbf{Quantum dot fabrication uniformity.} \textbf{a,} Room-temperature yield map of the targeted \SI{80}{\nano\meter} dot-to-dot-pitch design. Nine triple-dot devices per die were screened for gate-to-gate and gate-to-implant leakage. The pass criteria are defined in the main text. \textbf{b,} Capacitance equivalent thickness (CET) of the GL1, GL2, and GL3 gate oxides. The \textbf{inset} shows the measured dies on the wafer. \textbf{c,} Mean plunger-to-plunger spacing $d_\mathrm{pp}$ (Fig.~\ref{fig:m1}a) measured by CDSEM, averaged over four devices per die. The wafer-wide standard deviation is $\sigma=$ \SI{0.8}{\nano\meter}. \textbf{d,} Overlay error along [110] (Fig.~\ref{fig:m1}) between GL2 and GL3 measured after GL3 lithography. The wafer-wide standard deviation is $\sigma=$ \SI{1.5}{\nano\meter}. The gate etch further reduces the overlay error, with typical values of $\left|\mu\right|+3\sigma<$~\SI{3}{\nano\meter}~\cite{beyne_300mm_2025}.}
\end{figure}

\section*{Fabrication uniformity}
\noindent To assess the uniformity of the EUV-lithography fabrication process across the wafer, we monitored key process metrics relevant to the formation and operation of gate-defined quantum dots. Figure~\ref{fig:m2}a shows the room-temperature gate-to-gate leakage yield for the targeted \SI{80}{\nano\meter} dot-to-dot-pitch design. Leakage testing was performed at the device level by biasing each electrostatic gate to \SI{1}{\volt} while grounding all other gates. To protect the devices, we use a current compliance of \SI{10}{\nano\ampere}. To confirm that the probecard was well connected to the device, we verified whether both ends of the ESR antenna were shorted. The die under test was considered to pass if no gate in any of the nine tested triple-dot devices exhibited a gate-to-gate leakage current exceeding \SI{2}{\nano\ampere}. Electrical screening of all 89 dies on the \SI{300}{\milli \meter} wafer yielded a \SI{100}{\percent} pass rate.

\begin{figure*}[bt]
\includegraphics[width=\textwidth]{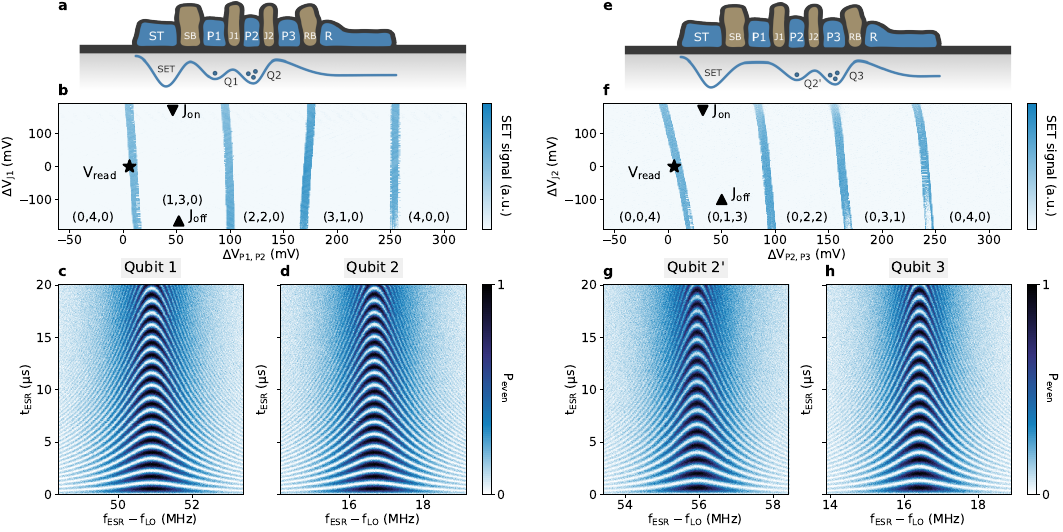}
\caption{\label{fig:m3}\textbf{Single-qubit operation of device A. a, e,} Cross-section of the triple-dot device illustrating the electron configuration under the plunger gates P1 and P2 (P2 and P3). \textbf{b, f,} Charge stability diagrams as a function of plunger-gate detuning $\Delta V_\mathrm{P1,P2}=\Delta V_\mathrm{P1}=-\Delta V_\mathrm{P2}$ ($\Delta V_\mathrm{P2,P3}=\Delta V_\mathrm{P2}=-\Delta V_\mathrm{P3}$) and exchange-gate voltage $\Delta V_\mathrm{J1}$ ($\Delta V_\mathrm{J2}$), showing four isolated electrons in the double-dot systems. Voltage operating points for single-qubit operation ($J_\mathrm{off} < \SI{1}{\kilo\hertz}$), two-qubit operation ($J_\mathrm{on} \approx \SI{2}{\mega\hertz}$), and readout are indicated by a triangle, inverted triangle, and star, respectively. Device B was also measured in the (1,3) charge configuration in both double-dot systems and showed similar charge stability diagrams. a.u., arbitrary units. \textbf{c, d, g, h,} Rabi chevrons of qubit~1 (\textbf{c}), qubit~2 (\textbf{d}), qubit~2$'$ (\textbf{g}), and qubit~3 (\textbf{h}). Real-time feedback is used to compensate for slow fluctuations of the Larmor frequency. The local oscillator frequency is set to $f_\mathrm{LO}=\SI{19.850}{\giga\hertz}$.}
\end{figure*}

On the same wafer, we measured the variation in the gate oxide thicknesses $t_\mathrm{ox}$ (Fig.~\ref{fig:m1}b) and the minimal spacing between adjacent plunger gates $d_\mathrm{pp}$ (Fig.~\ref{fig:m1}c) in selected dies. The capacitance-equivalent oxide thickness (CET) for the three gate layers is shown in Fig.~\ref{fig:m2}b. The thickness values were extracted from the gate-to-channel capacitance of a $W/L=\SI{5}{}/\SI{100}{\micro\meter}$ transistor, measured with the gate biased at \SI{2.5}{\volt} and the source and drain grounded. Measurements in nine dies distributed across the wafer show consistently low standard deviations for all gate layers. Furthermore, the gate oxides beneath GL2 and GL3 differ by less than \SI{0.5}{\nano\meter} on average, resulting in uniform gate lever arms for the plunger and barrier gates. The minimal plunger-to-plunger spacing $d_\mathrm{pp}$, measured on a subset of the devices subsequently screened for room-temperature leakage, exhibits a wafer-wide variation of less than \SI{1}{\nano\meter} (Fig.~\ref{fig:m2}c). Together with the \SI{1.5}{\nano\meter} wafer-scale standard deviation in GL2--GL3 overlay along the [110] direction of the dot array (Fig.~\ref{fig:m2}d), this enables the reproducible fabrication of sub-\SI{20}{\nano\meter} exchange gates across the wafer.

\section*{Qubit benchmarking}
\renewcommand{\arraystretch}{1.3}
\begin{table}[tb]
    \centering
    \caption{\textbf{Single-qubit performance.} Coherence times are extracted from Ramsey and Hahn echo measurements. We averaged over $N_\text{repeats}=20$ Ramsey measurements each integrated over $N_\text{shots}=$~1,000. The Hahn echo measurements are integrated over $N_\text{shots}=$~1,000. Rabi lifetimes are extracted from coherent oscillations of $f_\mathrm{Rabi}=\SI{640}{}$--\SI{860}{\mega \hertz} for a duration of $t_\mathrm{ESR}=$~\SI{100}{\micro\second}. In all experiments, we applied Larmor frequency feedback before each measurement shot~\cite{dumoulin_stuyck_silicon_2024}. On-target X-gate fidelities are obtained from gate set tomography (GST). Uncertainties correspond to 95\% confidence intervals.}
    \label{tab:t1}
    \begin{tabular}{clccccc}
    Device & Qubit & $T_2^*$ (\textmu s) & $T_2^\mathrm{Hahn}$ (\textmu s) & $T_2^\mathrm{Rabi}$ (\textmu s) & $X$ (\%) \\
    \hline
    \multirow{4}{*}{A} &  Q1  & \SI{9.3 \pm 0.3}{} & \SI{237 \pm 3}{} & \SI{244 \pm 119}{} & \SI{99.90 \pm 0.19}{} \\
                       &  Q2  & \SI{9.0 \pm 0.3}{} & \SI{244 \pm 4}{} & \SI{218 \pm 51}{} & \SI{99.69 \pm 0.15}{} \\
                       &  Q2$'$ & \SI{8.9 \pm 0.4}{} & \SI{231 \pm  4}{} & \SI{100 \pm 3}{} & \SI{99.95 \pm 0.22}{} \\
                       &  Q3  & \SI{9.3 \pm 0.4}{} & \SI{232 \pm  3}{} & \SI{182 \pm 30}{} & \SI{99.70 \pm 0.20}{} \\
    \hline                   
    \multirow{4}{*}{B} &  Q1  & \SI{10.1 \pm 0.5}{} & \SI{408 \pm 21}{} & \SI{174 \pm 10}{} & \SI{99.97 \pm 0.39}{} \\
                       &  Q2  & \SI{10.1 \pm 0.6}{} & \SI{388 \pm 21}{} & \SI{339 \pm 70}{} & \SI{99.73 \pm 0.31}{} \\
                       &  Q2$'$ & \SI{10.6 \pm 0.9}{} & \SI{417 \pm  7}{} & \SI{206 \pm 18}{} & \SI{99.91 \pm 0.67}{} \\
                       &  Q3  & \SI{10.7 \pm 0.8}{} & \SI{427 \pm  7}{} & \SI{330 \pm 82}{} & \SI{99.69 \pm 0.60}{} \\
   \hline
\end{tabular}
\end{table}

\begin{figure*}[bt]
\includegraphics[width=\textwidth]{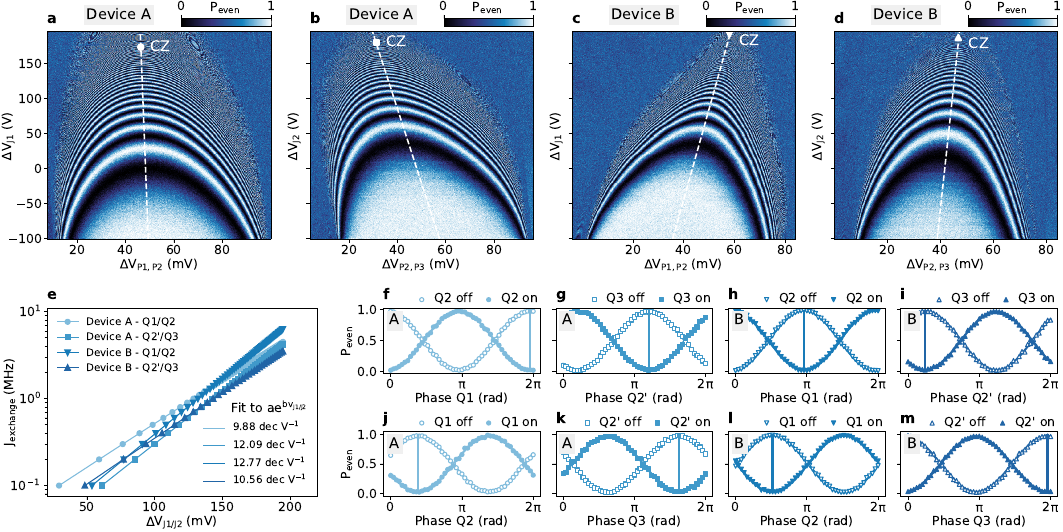}
\caption{\label{fig:m4}\textbf{Two-qubit operation. a--d,} Decoupled exchange oscillation fingerprints measured at a fixed exchange time $t_\mathrm{exchange}=10\,\unit{\micro\second}$, as a function of plunger detuning $\Delta V_\mathrm{P1,P2}$ ($\Delta V_\mathrm{P2,P3}$) and exchange gate voltage $\Delta V_\mathrm{J1}$ ($\Delta V_\mathrm{J2}$) for device A (\textbf{a,b}) and device B (\textbf{c,d}). The circle, square, triangle, and inverted triangle mark the operating points used for the CZ gates. \textbf{e,} Exchange interaction as a function of exchange gate voltage, extracted along the dashed lines in \textbf{a--d}. Averaging exponential fits across all four double-dot systems yields an exchange controllability of $11.3(2.6)$ dec V\textsuperscript{-1}. \textbf{f--m,} Calibration of the single-qubit phase correction for the CZ gate, obtained by preparing the target spin in a superposition, applying a CPHASE gate, and performing a virtual phase rotation, for qubit~1 (2$'$) (\textbf{f--i}) and qubit~2 (3) (\textbf{j--m}) in both devices. Vertical lines indicate the phase values at which the target qubit flips conditional on the control qubit being in the on/off state.}
\end{figure*}

\noindent 
To assess qubit quality, we performed in-depth characterization on two triple-dot devices (device A and B). The devices were operated in a $^3$He/$^4$He dilution refrigerator at a base temperature of \SI{10}{\milli\kelvin}. Four electrons were confined in the double quantum dot defined by plunger gates P1/P2, which we will refer to as qubits Q1 and Q2, and P2/P3 or qubits Q2$'$ and Q3, respectively, operated in isolated mode (compare Fig.~\ref{fig:m3}a,e). Tunnel coupling and exchange interaction were tuned via interstitial exchange gates (J1 and J2)~\cite{veldhorst_two-qubit_2015}. Electrons were loaded from a two-dimensional electron gas beneath a reservoir gate. Spin parity was read out via Pauli spin blockade (PSB) at the (1,3)–(0,4) charge transition using a d.c. SET (Fig.~\ref{fig:m3}b,f). To evaluate charge-noise, we performed SET noise spectroscopy, revealing a $A/f^\alpha$ power spectral density with $\alpha$=\SI{0.33\pm0.02}{}--\SI{0.37\pm0.02}{} and $A$=\SI{0.59\pm0.01}{}--\SI{0.65\pm0.02}{\micro \electronvolt / \sqrt{\hertz}} at $\SI{1}{\hertz}$ (Extended Data Fig.~\ref{fig:e2}), comparable to prior imec-fabricated devices~\cite{elsayed_low_2024,stuyck_demonstration_2024,steinacker_industry-compatible_2025}.

\renewcommand{\arraystretch}{1.3}
\begin{table*}[tb]
    \centering
    \caption{\textbf{Two-qubit performance.} Operation fidelities of all four double-dot systems extracted from GST experiments performed in the two-qubit space. Uncertainties correspond to 95\% confidence intervals.}
    \label{tab:t2}
    \begin{tabular*}{\textwidth}{@{\extracolsep{\fill}}ccccccccc}
    Device & Qubits & SPAM & II & XI & IX & ZI & IZ & CZ \\
    \hline
    \multirow{2}{*}{A} & Q1-Q2 & \SI{99.12 \pm 0.16}{} & \SI{99.62 \pm 0.12}{} & \SI{98.99 \pm 0.08}{} & \SI{98.68 \pm 0.11}{} & \SI{99.89 \pm 0.04}{} & \SI{99.64 \pm 0.06}{} & \SI{99.12 \pm 0.10}{} \\
                       & Q2$'$-Q3 & \SI{99.23 \pm 0.13}{} & \SI{99.50 \pm 0.12}{} & \SI{98.82 \pm 0.08}{} & \SI{98.35 \pm 0.07}{} & \SI{99.81 \pm 0.06}{} & \SI{99.72 \pm 0.06}{} & \SI{99.06 \pm 0.08}{}\\
    \hline                   
    \multirow{2}{*}{B} 
                       & Q1-Q2 & \SI{99.82 \pm 0.24}{} & \SI{99.76 \pm 0.18}{} & \SI{98.99 \pm 0.14}{} & \SI{98.87 \pm 0.12}{} & \SI{99.85 \pm 0.11}{} & \SI{99.80 \pm 0.11}{} & \SI{98.82 \pm 0.15}{} \\
                       & Q2$'$-Q3 & \SI{99.19 \pm 0.41}{} & \SI{99.41 \pm 0.40}{} & \SI{99.06 \pm 0.22}{} & \SI{98.67 \pm 0.22}{} & \SI{99.79 \pm 0.19}{} & \SI{99.69 \pm 0.21}{} & \SI{98.20 \pm 0.22}{} \\
    \hline
\end{tabular*}
\end{table*}

An external magnetic field $B_0$=\SI{0.7}{\tesla} pointing along [110] defines the spin qubit basis. Single-qubit control is achieved using on-resonance microwave excitation applied to an aluminum electron-spin resonance (ESR) antenna located approximately \SI{150}{\nano \meter} above the dot array (see Fig.~\ref{fig:m1}b), generating an in-plane oscillating field $B_1$ perpendicular to $B_0$. Rabi chevrons are measured by sweeping the pulse duration $t_\text{ESR}$ and frequency $f_\text{ESR}$ at exchange gate voltages corresponding to minimal residual exchange coupling $\Delta V_\text{J} = J_\text{off}$ (see Fig.~\ref{fig:m3}b--d,f--h). Rotations about the $\hat{x}$-axis, or X$_{\pi/2}$ gates, are implemented with resonant pulses of calibrated duration, while $\hat{z}$-axis rotations, or Z$_{\pi/2}$ gates, are realized virtually via microwave signal phase changes~\cite{vandersypen_nmr_2005}. Single-qubit metrics for both devices and corresponding double-dot systems are summarized in Table~\ref{tab:t1}. For two-qubit control, exchange is turned on by pulsing to $\Delta V_\text{J} = J_\text{on}$ for a calibrated duration $t_\text{CZ}$ (Fig.~\ref{fig:m3}b,f). Simultaneously, the plunger gates are pulsed to the symmetric operation point to minimize sensitivity to detuning noise (Fig.~\ref{fig:m4}a--d)~\cite{reed_reduced_2016}. Along the operating axis that minimizes sensitivity to charge-noise fluctuations, the exchange frequency is fitted to extract an exchange controllability of $11.3(2.6)$~\SI{}{dec}V$^{-1}$, averaged across devices and two-qubit systems (Fig.~\ref{fig:m4}e). Compared to previous work on e-beam-defined SiMOS devices~\cite{steinacker_industry-compatible_2025}, this reduced variation in exchange controllability observed here reflects the reproducibility of oxide thickness, plunger-to-plunger spacing, and overlay alignment enabled by the EUV lithography.

The pulsed exchange interaction is combined with single-qubit phase rotations to compensate the accumulated unconditional phase, thereby implementing a controlled-phase (CZ) gate~\cite{veldhorst_two-qubit_2015} (Fig.~\ref{fig:m4}e--l). During operation, real-time feedback is used to stabilize the SET operating point and correct for drift in the single-qubit Larmor and Rabi frequencies~\cite{dumoulin_stuyck_silicon_2024}. In addition, a heralded initialization protocol verifies preparation in the $\ket{\downarrow \downarrow}$ state and repeats initialization when necessary~\cite{huang_high-fidelity_2024,steinacker_bell_2025}.

To benchmark qubit performance, we perform GST comprised of the gate set \{II, XI, IX, ZI, IZ, CZ\}, where XI (IX) denotes a $\pi/2$ rotation about the $\hat{x}$-axis on the first (second) qubit, during which the second (first) qubit remains idle. Analogously, ZI (IZ) denotes the corresponding $\pi/2$ $\hat{z}$-rotation. Single-qubit X rotations are implemented using square pulses of \SI{300}{}--\SI{400}{\nano \second} at the qubit Larmor frequency $f_\text{L}$. Virtual Z rotations are realized via microwave signal phase changes, with a latency of $\leq$\SI{100}{\nano \second} set by the FPGA. The CZ gate is implemented by pulsing the exchange gate J for a duration $t_\text{CZ}$=\SI{120}{}--\SI{140}{\nano \second}. For comparison, the idle gate duration is matched to $t_\text{CZ}$. Germ circuits up to length 16 are used, resulting in 12,263 unique sequences~\cite{nielsen_probing_2020}. GST resolves a detailed error taxonomy and typically yields more stringent fidelity estimates than other quantum characterization, verification and validation (QCVV) methods~\cite{tanttu_assessment_2024,blume-kohout_quantum_2025,hashim_practical_2025}.

The performance of the four two-qubit systems is summarized in Table~\ref{tab:t2}. All operations, including combined state preparation and measurement (SPAM) fidelity, approach or exceed the threshold required for surface-code error correction. The extracted fidelities and error generators are comparable to previous results from devices fabricated using e-beam lithography~\cite{stuyck_demonstration_2024,steinacker_industry-compatible_2025}, indicating that the EUV fabrication process does not introduce additional error mechanisms. In the two-qubit space, dephasing of the idling qubit (stochastic IZ and ZI terms) remains the dominant contribution to the gate infidelity (Extended Data Fig.~\ref{fig:e3}).

\section*{Discussion} 
\noindent
A key challenge in scaling silicon spin-qubit processors beyond small arrays is achieving sufficiently uniform electrostatic confinement and coupling across many qubits to enable reproducible high-fidelity operation~\cite{borsoi_shared_2024,tosato_crossbar_2026,li_tri-linear_2026}. Variations in gate geometry and placement directly affect the confinement potential, tunnel coupling, and exchange interaction, increasing calibration overhead and complicating large-scale control. Across all four double-dot systems, we observe similar operating conditions, reproducible exchange turn-on, and high-fidelity qubit control. Together with the wafer-scale uniformity of the EUV-defined triple-dot structures and gate-stack dimensions, these results demonstrate a high degree of fabrication-process reproducibility. This reduction in fabrication-induced variability is particularly important for gate-defined spin qubits, which, unlike several competing qubit modalities, must first be electrostatically tuned into the desired charge configuration and operating regime before qubit operation can be established. By reducing device-to-device variations in the electrostatic landscape, EUV lithography increases the likelihood that common gate-voltage settings produce consistent confinement potentials, tunnel couplings, and operating points across qubits. This predictability reduces calibration overhead, simplifies automated tune-up, and supports scalable control architectures.

\section*{Conclusion} 
\noindent
This work demonstrates that advanced optical lithography can satisfy the stringent dimensional requirements of SiMOS quantum-dot architectures while providing the reproducibility required for scalable operation. Beyond the qubit performance demonstrated here, fabrication in a \SI{300}{\milli\meter} semiconductor pilot line enables the systematic study of large device ensembles, allowing variability, yield, and performance distributions to be quantified at a scale that is difficult to perform in laboratory-scale fabrication environments. Together, these capabilities establish a foundation for the development of increasingly uniform, manufacturable, and scalable silicon quantum processors.

Looking ahead, continued advances in semiconductor lithography may enable increasingly compact quantum-dot architectures. The sub-\SI{20}{\nano\meter} dimensions demonstrated here were achieved using 0.33 NA EUV lithography and a three-layer overlapping-gate design. Future generations of EUV patterning, including 0.55 NA EUV, may provide access to substantially smaller critical dimensions in SiMOS devices, opening a route towards simplified single-layer gate architectures with higher integration density and reduced fabrication complexity~\cite{beyne_300mm_2025}. By reducing the architectural overhead associated with multi-layer gate stacks, such developments could facilitate the integration of larger and more densely packed qubit arrays.

The present work also highlights a growing mismatch between advances in quantum-device manufacturing and the throughput of cryogenic characterization. Modern semiconductor fabrication facilities can produce thousands of candidate quantum devices on a single wafer, yet comprehensive qubit characterization remains largely confined to individual dilution refrigerators, as reflected in this study. As fabrication uniformity and yield improve, the ability to identify, benchmark, and compare large numbers of devices efficiently will become increasingly important. Cryogenic wafer-probing platforms are beginning to enable high-throughput measurements of charge transport and basic device characteristics~\cite{neyens_probing_2024}. However, extending such approaches to full qubit characterization will require capabilities typically available only in dedicated quantum-measurement systems, including lower temperatures, large magnetic fields, and high-frequency control. In parallel, multiplexed cryogenic measurement platforms have demonstrated the characterization of increasing numbers of quantum-dot devices within a single dilution refrigerator~\cite{thomas_rapid_2025,candido_investigation_2025}. Further advances in scalable cryogenic test infrastructure will therefore be essential for translating advances in quantum-device manufacturing into larger and more reliable silicon quantum processors.

\bibliography{mybib}

\section*{Methods}
\noindent \textbf{Measurement.} 
Both devices were measured in a Bluefors LD400 dilution refrigerator and mounted back-to-back on the cold finger. An American Magnetics AMI430 supplied the static external magnetic field to both devices along the [110] direction of the Si lattice.  Microwave pulses generated by a Keysight PSG8267D Vector Signal Generator applied to the antenna above the dot array supplied a driving magnetic field along the [$\overline{1}$10] direction. In-phase and quadrature (I/Q) and pulse modulation waveforms are generated by a Quantum Machines (QM) Operator-X+ (OPX+).

A Q-Devil QDAC-II supplied the d.c. voltages through filtered lines with a bandwidth of up to $\SI{20}{\hertz}$. The OPX+, a field-programmable gate array (FPGA), generated the dynamic voltage pulses that were combined with the d.c. biases using custom linear bias combiners at room temperature. The dynamic pulse lines in the fridge have a bandwidth of up to \SI{50}{\mega\hertz}, which translates into a minimum rise time of $\SI{20}{\nano\second}$. The OPX+ has a sampling time of $\SI{1}{\nano\second}$.

The charge sensor comprises a single-island SET in d.c. mode. The SET current is amplified using a room-temperature I–V converter (Basel SP983c) and sampled by a QM OPX+. The wait and integration time per shot is kept constant at $t_\text{read} = t_\text{wait} + t_\text{int} = \SI{200}{\micro \second}$.

We optimized gate times based on a trade-off between speed and lifetime. The magnetic field amplitude $B_{0} = \SI{0.7}{\tesla}$ was chosen to maximize the single-qubit Rabi Q-factor $Q_\text{Rabi} = f_\text{Rabi}T_{2}^\text{Rabi}$. Simultaneously, CZ gates require the exchange interaction to be smaller than the Zeeman energy difference of the two qubits ($J_\text{exc} \ll \Delta E_\text{Z}$). The requirement of fast two-qubit gates determined the chosen $B_{0}$ values.

\section*{Data Availability}
\noindent The data supporting this work will be made available in a Zenodo repository upon publication.

\section*{Acknowledgments}
\noindent We thank Patrick Carolan for performing the TEM analysis. We thank Fay E. Hudson and Kok Wai Chan for support in device packaging. We thank Andrii Torgovkin and Sophia Wolczak for assistance with the cryogenic setup. We thank Jesus Cifuentes Pardo and Santiago Serrano Ramirez for useful discussions and input. This work was supported by the imec Industrial Affiliation Program on Quantum Computing and the European Union’s Horizon 2020 Research and Innovation Program under grant agreement No. 101174557 (QLSI2). T. V. C. acknowledges the support of the Research Foundation Flanders through the Strategic Basic Research PhD program (grant No. 1SAAI26N). N. D. S. is the recipient of an Australian Research Council Industrial Fellowship (project No. IE240100252) funded by the Australian Government.

\subsection*{Author Contributions}
\noindent C. G. designed the devices. S. Beyne, S. Kubicek, J. D. B, Y. H., S. S., S. Kaushik, Y. J., Y. S., and R. L. fabricated the devices. P. S. and T. V. C. conducted the experiments with W. H. L.'s and N. D. S.'s supervision as well as input from B. R., T. T., C. C. E., C. H. Y., D. W., A. S., A. L., A. S. D., and K. D. G.. F. K. U. packaged the devices. S. Baudot performed room-temperature screening. A. D. assisted with the cryogenic screening. E. V. assisted with the experimental setup. T. V. C. and P. S. wrote the manuscript, with input from all authors. K. D. G., A. S. D., D. W., N. D. S., W. H. L., and B. R. supervised the project.

\section*{Competing Interests}
\noindent A. S. D. is CEO and a director of Diraq Pty Ltd. E. V., T. T., C. C. E., C. H. Y., A. S., A. L., W. H. L., N. D. S, and A. S. D. declare equity interest in Diraq. Other authors declare no competing interest.

\newpage
\setcounter{figure}{0}
\setcounter{table}{0}
\captionsetup[figure]{name={\bf{Extended Data Fig.}},labelsep=line,justification=RaggedRight,font=small,singlelinecheck=false}

\begin{figure*}[ht]
\includegraphics[width=1\textwidth,angle = 0]{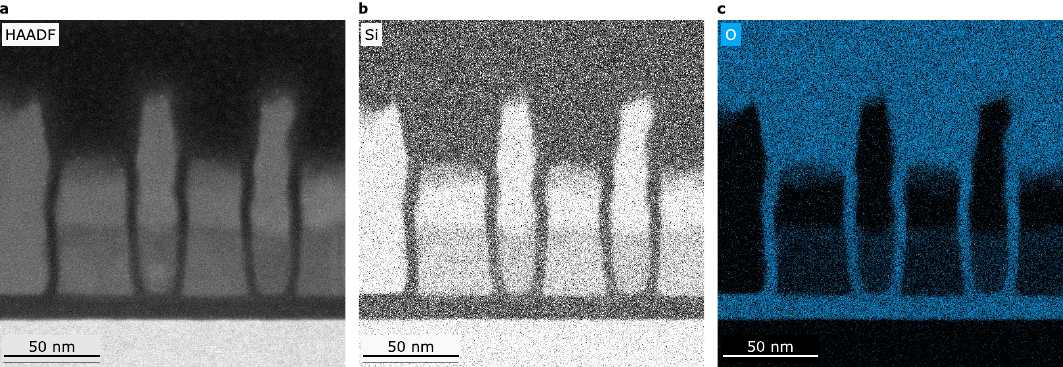}
\caption{\label{fig:e1}\textbf{Energy dispersive spectroscopy (EDS) of the gate stack.} \textbf{a,} close-up HAADF-STEM image of the cross-section in Fig.~\ref{fig:m1}b of the main text. \textbf{b, c,} EDS images of the same cross-section as \textbf{a}, depicting the silicon (\textbf{b}) and oxygen (\textbf{c}) atom distributions. The silicon substrate and polysilicon gates are visible in \textbf{b}. The SiO$_2$ layers that separate the gates from the substrate and each other are clearly visible in blue in \textbf{c}. The lower part of the gates in \textbf{b, c} are slightly brighter due to the oxide that is present before and/or behind the polysilicon gates.}
\end{figure*}

\begin{figure*}[ht]
\includegraphics[width=1\textwidth,angle = 0]{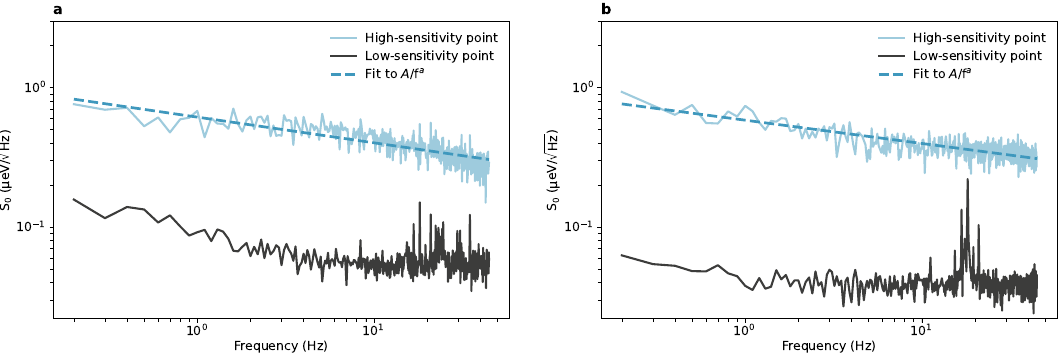}
\caption{\label{fig:e2}\textbf{Single electron transistor charge noise spectrum.} \textbf{a, b,} Charge noise spectrum at high and low-sensitivity points of a Coulomb oscillation for device A and B, respectively. We fast Fourier transform the SET current measurements to a voltage noise spectrum, divide it by the slope of the Coulomb peak, and convert it using the lever arm extracted from a Coulomb diamond measurement. The spectra are averaged over five measurements performed along the Coulomb peak. The charge noise at $\SI{1}{\hertz}$ is $A=\SI{0.62\pm0.02}{\micro \electronvolt / \sqrt{\hertz}}$ and $A=\SI{0.59\pm0.01}{\micro \electronvolt / \sqrt{\hertz}}$. The dashed line is a power law fit of the power spectral density with exponents of $a=\SI{0.37\pm0.02}{}$ and $a = \SI{0.33\pm0.02}{}$.}
\end{figure*}

\begin{figure*}[ht]
\includegraphics[width=1\textwidth,angle = 0]{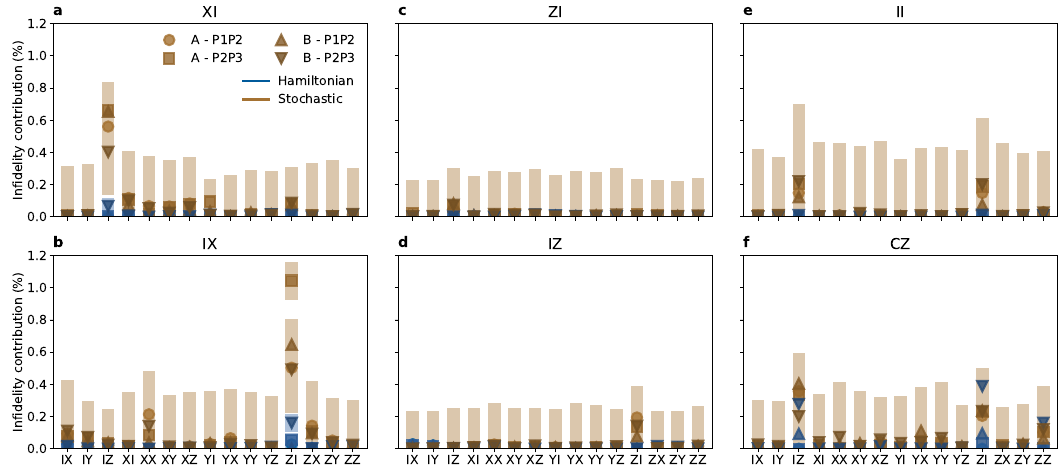}
\caption{\label{fig:e3}\textbf{GST breakdown. a--f} Infidelity contributions of Hamiltonian and stochastic error generators across both devices (A \& B) and double-dot systems (P1P2 \& P2P3) for the XI (a), IX (b), ZI (c), IZ (d), II (e) and CZ (f) gates from GST. Hamiltonian errors contribute to the infidelity in second order, whereas stochastic errors contribute in first order. In the two-qubit context, the infidelity is dominated by the stochastic ZI and IZ error generator components, originating from the idling qubit(s). Error bars represent the 95\% confidence level.}
\end{figure*}

\end{document}